# A new indicator for the AMOC strength still gives no indication of an imminent collapse

Erhard Reschenhofer[1]

**Abstract**

The results of a recent simulation with a complex global climate model suggest that the overturning component of the freshwater transport at the southern boundary of the Atlantic could be used as an early-warning indicator of an AMOC collapse. However, there are two shortcomings. Firstly, the simulation is based on some implausible assumptions. It is therefore not clear whether this new indicator will still be helpful in a more realistic setting. Secondly, the statistical methods employed in the simulation work only for time series that are much longer than the currently available historical series. As not much can be done in the short term about the first issue, this paper focuses on the second issue. It is shown that it is possible to use alternative statistical methods that can do the job at least as well and, moreover, are also suitable for short series. Applying these more appropriate methods to historical data obtained with an ocean reanalysis system, no indication of an imminent AMOC collapse is found.

## 1. Introduction

For their prediction of the imminent collapse of the Atlantic meridional overturning circulation (AMOC), Ditlevsen and Ditlevsen (2023) used the increase in the (first-order) autocorrelation of a sea surface temperature (SST)-based AMOC proxy as an early-warning sign. Arguing that many assumptions are required for this prediction (some of which are purely statistical) and the AMOC proxy may also not represent the AMOC behavior adequately (for a more detailed criticism see Reschenhofer, 2023, Reschenhofer, 2024a, 2024b), van Westen et al. (2024) emphasized the strong need for a more reliable early-warning sign. Because of the importance of the freshwater balance of the Atlantic for the stability of the AMOC (Rahmstorf, 1996; de Vries and Weber, 2005; Jackson, 2013), the net meridional freshwater transport by the overturning circulation at the southern boundary (34°S) of the Atlantic ($F_{ovS}$) could be a helpful indicator. Indeed, conducting a simulation with a complex global climate model, van Westen et al. (2024) found that the indicator $F_{ovS}$ reaches its minimum about 10 to 40 years before the tipping point of the AMOC is reached. They therefore proposed to linearly extrapolate the time derivative of a smoothed version of $F_{ovS}$ (obtained by fitting cubic splines to 50-year averages) in order to find the point in time where it goes through zero. The collapse is expected to occur about 25 years after that point. In addition, van Westen et al. (2024) also looked at classical early-warning signs such as an increase in the variance and the autocorrelation of some variable of interest, for example the indicator $F_{ovS}$ or an SST-based AMOC proxy. In this case, they forecasted the time of the collapse by estimating the autocorrelation with rolling windows and extrapolating the increasing autocorrelation until it reaches the value 1 (for details see Supplementary Materials of Westen et al., 2024, as well as Ditlevsen and Ditlevsen, 2023).

---

[1] *Retired from University of Vienna, E-mail: erhard.reschenhofer@univie.ac.at*

The simulation results regarding the performance of the two forecasting methods are mixed and depend, of course, on the respective method and the respective indicator, but also on the choice of the estimation period and the smoothing method (for the reduction of the high variability in the data). A further uncertainty factor is that the simulation results were obtained under quite unrealistic assumptions. In their simulation, Westen et al. (2024) found a gradual decrease in the AMOC strength under increasing freshwater forcing and eventually a much steeper decline, which may pass for a collapse, although it extends over an entire century. This proof of the possible existence of an AMOC collapse in a complex global climate model, which takes the entire climate system into account, is certainly a success. However, one would expect that a more complex model would also allow the choice of a more realistic setting, e.g., a realistic amount of meltwater inflow into the North Atlantic caused by global warming over a reasonable period of 100 or 200 years. In contrast, van Westen et al. (2024) increased the freshwater input in their simulation over a period of more than 2000 years and kept increasing even when the temperature in Northern Europe was already falling dramatically (and the Greenland Ice Sheet was growing!), which makes it difficult to assess the relevance of the simulation results for the real world. This is especially true for individual results such as a temperature drop of about 35°C for Bergen (Norway) in February, which admittedly makes good fodder for the media.

In a real-world forecasting situation, historical time series must be used, which are not very reliable because they are usually very short and consist of either estimates or proxies. In the case of the indicator $F_{ovS}$, estimates derived from different reanalysis and assimilation products show huge differences which implies that the forecast accuracy will critically depend on the choice of the product. Also, the reanalysis periods are relatively short considering the high variability (GLORYS12V1: 1993-2020, SODA.15.2: 1980-2020, ORAS5: 1958-2022, ORA-20C: 1900-2009, ECCO-V4r4: 1992-2017). The longest period covers 110 years (for the product ORA-20C), the shortest only 26 years. However, the authors' worry that 200 years might be needed for a reliable forecast is certainly too pessimistic. A significant improvement can already be achieved just by using the simple Hodrick-Prescott (HP) filter for smoothing instead of fiddling around with 50-year averages. Moreover, it is not the total number of years that matters, but only the few years immediately before and after the event of interest occurs. For this reason, the shorter period from 1958 to 2022 (ORAS5) is actually more helpful than the longer period from 1900 to 2009 (ORA-20C). Since there is no obvious decline before the late 1970s, this early period is absolutely useless for the identification of the minimum. Only when a downward trend emerges can we look for signs of a slowdown.

Clearly, using window sizes of 50 or 70 years is not an option when we are dealing with time series that only span 65 (ORAS5), 110 (ORA-20C), or 152 years (AMOC proxy of Ditlevsen and Ditlevsen, 2023). In the next section, it will be shown how it is still possible to analyze such short time series in a meaningful way. For illustration, historical $F_{ovS}$ data obtained with an ocean reanalysis system of the European Centre for Medium-Range Weather Forecasts (ECMWF) will be used (Zuo et al., 2019). For the statistical analysis, the free statistical software R (R Core Team, 2022) will be used. Section 3 concludes.

## 2. Assessing the usefulness of the indicator $F_{ovS}$

The Ocean Reanalysis System 5 (ORAS5) of the ECMWF provides estimates of the historical state of the global ocean from 1979 until today. A longer reanalysis period can be obtained by using a backward extension of ORAS5, which is available from 1958 onwards (Zuo et al., 2019). The historical $F_{ovS}$ from 1958 to 2022 derived from the product ORAS5 (https://doi.org/10.24381/cds.67e8eeb7) was kindly provided by van Westen et al. (2024) online at https://doi.org/10.5281/zenodo.10461549. Figure 1.a shows this time series together with an estimate of its trend obtained with the Hodrick-Prescott (HP) filter (using the R function hpfilter of the package mFilter and choosing the value 2000 for the tuning parameter $\lambda$). The HP filter estimates the trend $F_1,...,F_n$ of a time series $X_1,...,X_n$ by minimizing

$$\sum_{t=1}^{n}(X_t - F_t)^2 + \lambda \sum_{t=3}^{n}((F_t - F_{t-1}) - (F_{t-1} - F_{t-2}))^2, \tag{1}$$

where the tuning parameter $\lambda$ determines the degree of smoothing. There is not the slightest indication that the trend is heading for a minimum (see Figure 1.a) or that the differenced trend approaches zero (see Figure 1.b). Choosing the smaller value $\lambda = 200$ results in the somewhat too volatile trend shown in Figure 1.e, which entails a higher risk of exhibiting "false" minima. Indeed, the narrow minimum in 2001 (see Figures 1.e and 1.f) could have led us to predict an AMOC collapse 25 years later (i.e., already in 2026!).

Rolling windows (used by Boers, 2021; Ditlevsen and Ditlevsen, 2023; van Westen et al., 2024) are not indispensable when assessing the changes in second moments. Cumulative plots are a good alternative (see Reschenhofer, 2023, Reschenhofer, 2024a, 2024b). In the case of the variance, the procedure is obvious. The squared deviations from the trend

$$U_t^2 = (X_t - F_t)^2 \tag{2}$$

are plotted cumulatively. Figure 1.c shows that there is a structural break in the late 1990s. The constant slope in the first regime means that the variance is constant in this period. The same is true for the second regime. Obviously, the latter variance is greater than the former because the incline is steeper in the second regime. There is no indication of a further increase in the variance. If there were a further change at all, it would only be downwards. However, it is too early to say whether the possible second structural break at about 2010 is real or not.

In the case of the autocorrelation $\rho$, things are a little more complicated. The key difference is not that only one value is needed to estimate the variance when using statistic (2) and two values are needed to estimate the autocorrelation when using the statistic

$$\tilde{\rho}_t = \frac{2U_t U_{t-1}}{U_{t-1}^2 + U_t^2} \tag{3}$$

(Burg, 1967, 1975). The decisive disadvantage of the latter statistic is rather its large bias. This bias could be largely avoided under certain circumstances if the statistic

$$R_t = \frac{\pi}{\pi-2} sign(U_{t-1}U_t) \frac{\min(|U_{t-1}|,|U_t|)}{\max(|U_{t-1}|,|U_t|)} \tag{4}$$

is used instead (see Reschenhofer, 2017a, 2017b, 2019). But since the necessary prerequisite of a small autocorrelation ($|\rho| < 0.2$) is not fulfilled in the case of the indicator $F_{ovS}$, there needs to be another alternative. The unexpected solution is to replace the ratio (3) by a difference (see Reschenhofer, 2024b). Indeed, monitoring the statistic

$$c\, U_t^2 - U_t U_{t-1} \tag{5}$$

for various values of $c$ allows the assessment of the size of the autocorrelation at any point of time. A decrease/increase in (5) suggests that the autocorrelation is grater/less than $c$. In Figure 1.d, the statistic (5) is plotted cumulatively over time for $c = 0.1, 0.3, 0.5$. Based on this graphic, it is safe to conclude that the autocorrelation is always between 0.1 and 0.5. There is absolutely no indication that the autocorrelation could approach 1 in the foreseeable future. Hence, there is no reason at this point to worry about the extrapolation. Clearly, a simple linear extrapolation is not the only option. Ditlevsen and Ditlevsen (2023) and Reschenhofer (2024a) also used linear extrapolation for their predictions but carried out a specific nonlinear transformation beforehand that was motivated by the underlying dynamic model.

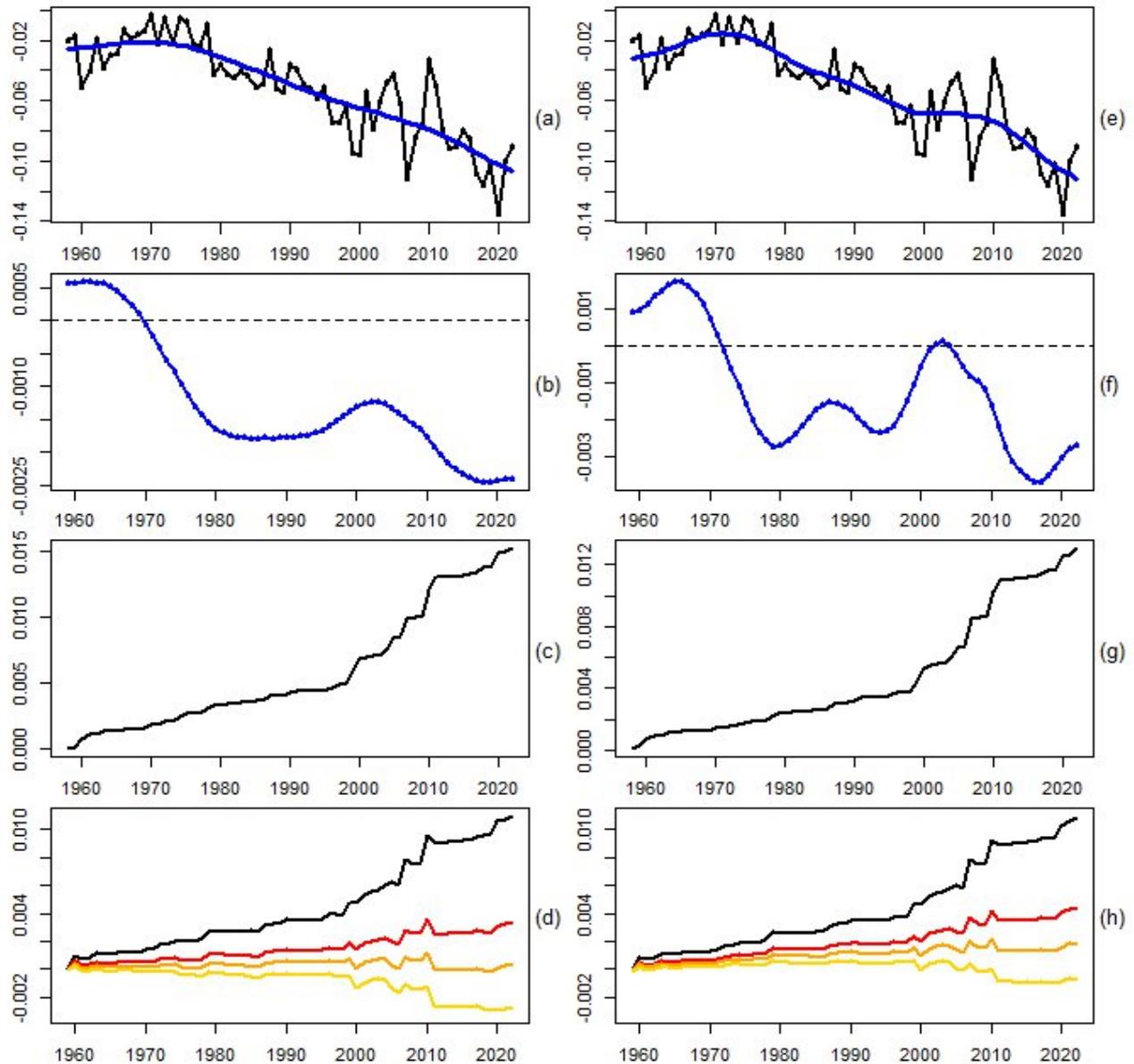

Figure 1: Analysis of the historical $F_{ovS}$ from 1958 to 2022 derived from the product ORAS5 (https://doi.org/10.24381/cds.67e8eeb7, https://doi.org/10.5281/zenodo.10461549)

(a), (e): Hodrick-Prescott trend (blue) obtained with $\lambda = 2000$ (a) and $\lambda = 200$ (e)

(b), (f): First differences of trend

(c), (g): Cumulative sum of squared trend residuals: $U_1^2, U_1^2 + U_2^2, U_1^2 + U_2^2 + U_3^2, \ldots$

(d), (h): Cumulative sums of differences $c\,U_t^2 - U_t U_{t-1}$ with $c = 1$ (black), $c = 0.5$ (red), $c = 0.3$ (orange), and $c = 0.1$ (yellow).

## 3. Discussion

Shortly after the first critical responses (Reschenhofer, 2023, Reschenhofer, 2024a, 2024b) to the prediction of an AMOC collapse around mid-century (Ditlevsen and Ditlevsen, 2023) were published, another worrying study appeared. Van Westen et al. (2024) found indications that "the AMOC is on course to tipping". However, the new study has similar weaknesses as the old one. For the derivation of the respective early-warning signs, both the dynamic model of Ditlevsen and Ditlevsen (2023) and the simulation of van Westen et al. (2024) used assumptions that are implausible or at best unverifiable. Moreover, the statistical methods proposed for the utilization of the early-warning signs (for example with regard to forecasting) are inadequate.

In particular, the use of estimation windows spanning several decades is inadmissible if the available time series are relatively short and there are serious indications of structural breaks. To illustrate, consider the simple case where there is only one structural break and all the 50 values in the first regime are small and all the 50 values in the second regime are large. Then the average of the first 50 values will be based on 50 small values, the next average on 49 small values and one large value, ..., and the last average on 50 large values. The steady increase visible in the plot of the moving averages would absurdly suggest creating forecasts with the help of a simple extrapolation.

Apart from the limited quantity of the available historical data, the quality is also a problem. Mostly there are only estimates or proxies. To get a feeling for the dependence of the results on the choice of the dataset, the analysis of Section 2 is repeated with different data. The product ORAS5 is replaced by the product ORA-20C. Figure 2 shows in accordance with Figure 1 once again that the downward trend is not slowing down and the autocorrelation is well below 0.5. The autocorrelation is also low in comparison to previous studies, which can easily be explained by the use of annual data instead of monthly data. But what matters most is that there is no recent increase in the autocorrelation.

As a cautious concluding statement, the following can be said. Assuming that the proposed early-warning signs are useful in some way and applying more appropriate statistical methods to the short series of available historical data, no indications of an imminent collapse of the AMOC can be found. This fruitless search includes in particular also the minimum of the indicator $F_{ovS}$ in combination with its variance increase, which is according to van Westen et al. (2024) "a very promising early warning signal for a (future) AMOC collapse". Of course, this does not mean that a collapse is not imminent, but merely that the evidence presented so far in favor of an imminent collapse is not reliable.

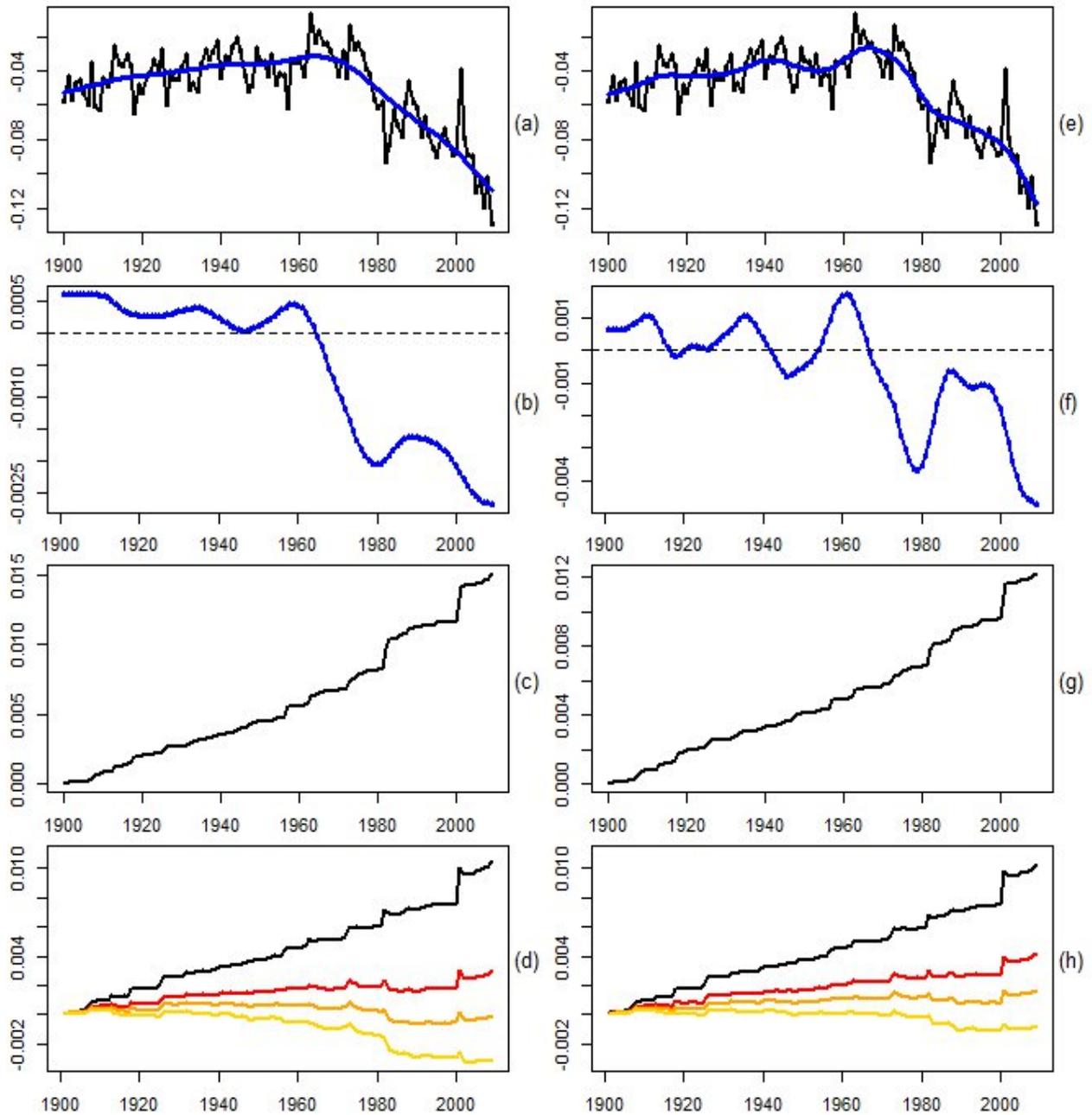

Figure 2: Analysis of the historical $F_{ovS}$ (1900-2009) derived from the product ORA-20C
(https://icdc.cen.uni-hamburg.de/thredds/catalog/ftpthredds/EASYInit/ora20c/opa0/catalog.html,
https://doi.org/10.5281/zenodo.10461549)

(a), (e): Hodrick-Prescott trend (blue) obtained with $\lambda = 2000$ (a) and $\lambda = 200$ (e)

(b), (f): First differences of trend

(c), (g): Cumulative sum of squared trend residuals: $U_1^2, U_1^2 + U_2^2, U_1^2 + U_2^2 + U_3^2, \dots$

(d), (h): Cumulative sums of differences $c\, U_t^2 - U_t U_{t-1}$ with $c = 1$ (black), $c = 0.5$ (red), $c = 0.3$ (orange), and $c = 0.1$ (yellow).